\documentclass[11pt]{article}
\usepackage[utf8]{inputenc}
\usepackage{amsfonts,epsfig}
\usepackage[hyphens]{url}
\usepackage{hyperref}
\usepackage{breakurl}
\usepackage{comment}
\usepackage{color}
\usepackage{bbm}

\usepackage{relsize}
\usepackage{fancyvrb}
\usepackage{amssymb,amsmath}
\usepackage{geometry}
\usepackage{sectsty}
\usepackage{caption}
\usepackage{subcaption}
\usepackage[normalem]{ulem}
\sectionfont{\sffamily\bfseries\upshape\large}
\subsectionfont{\sffamily\bfseries\upshape\normalsize}
\subsubsectionfont{\sffamily\mdseries\upshape\normalsize}
\makeatletter
\renewcommand\@seccntformat[1]{\csname the#1\endcsname.\quad}
\makeatother\renewcommand{\bibitem}{\vskip 2pt\par\hangindent\parindent\hskip-\parindent}

\makeatletter
\def\@maketitle{%
  \begin{center}%
  \let \footnote \thanks
    {\large \@title \par}%
    {\normalsize
      \begin{tabular}[t]{c}%
        \@author
      \end{tabular}\par}%
    {\small \@date}%
  \end{center}%
}
\makeatother

\title{\bf The prior can generally only be understood in the context of the likelihood\footnote{We thank Matt Hoffman for helpful comments and the National Science Foundation, Office of Naval Research, Institute for Education Sciences, and Sloan Foundation for partial support of this work.}\vspace{.1in}}
\author{Andrew Gelman\footnote{Department of Statistics and Department of Political Science, Columbia University.} \and Daniel Simpson\footnote{Department of Statistical Sciences, University of Toronto.} \and Michael Betancourt\footnote{Institute for Social and Economic Research and Policy, Columbia University.}\vspace{.1in}}
\date{28 Aug 2017\vspace{-.1in}}
\begin{document}\sloppy
\maketitle
\thispagestyle{empty}

\begin{abstract}

A key sticking point of Bayesian analysis is the choice of prior distribution, and there is a vast literature on potential defaults including uniform priors, Jeffreys' priors, reference priors,
maximum entropy priors, and weakly informative priors.  These methods,
however, often manifest a key conceptual tension in prior modeling:
a model encoding true prior information should be chosen without reference to the
model of the measurement process, but almost all common prior modeling
techniques are implicitly motivated by a reference likelihood.  In this
paper we resolve this apparent paradox by placing the choice of prior
into the context of the entire Bayesian analysis, from inference to 
prediction to model evaluation.

\end{abstract}

\section{The role of the prior distribution in a Bayesian analysis}

Both in theory and in practice, the prior distribution can play many
roles in a Bayesian analysis.  Perhaps most formally the prior serves
to encode information germane to the problem being analyzed, but in
practice it often becomes a means of stabilizing inferences in complex,
high-dimensional problems.  In other settings it is treated as little
more than a nuisance, serving simply as a catalyst for the expression 
of uncertainty via Bayes' theorem.

These different roles often motivate a distinction between ``subjective''
and ``objective'' choices of priors, but we are unconvinced of the 
relevance of this distinction (Gelman and Hennig, 2017).  We prefer to 
characterize Bayesian priors, and statistical models more  generally, 
based on the information they include rather than the philosophical
interpretation of that information.

The ultimate significance of this information, and hence the prior 
itself, depends on exactly how that information manifests in the final 
analysis.  Consequently the influence of the prior can only be judged 
within the context of the likelihood.

In the present paper we address an apparent paradox: Logically, the prior distribution should come before the data model, but in practice, priors are often chosen with reference to a likelihood function.

We resolve this puzzle in two ways, first with a robustness argument, recognizing that our models are only approximate, and in particular the relevance to any given data analysis of particular assumptions in the prior distribution depends on the likelihood; and, second, by considering the different roles that the prior plays in different Bayesian analyses.

\subsection{The practical consequences of a prior can depend on the data}

One might say that what makes a prior a prior, rather than simply a 
probability distribution, is that it is destined to be paired with a likelihood.  That is, the 
Bayesian formalism requires that a prior distribution be updated into
a posterior distribution based on new data.  

The practical utility of a prior distribution within a given analysis 
then depends critically on both how it interacts with the assumed probability model for the data in the context of the actual data that are
observed.  Consider, for example, a simple binomial 
likelihood with $n=75$ trials and some prior on the success probability, 
$p$.  If you observe $y=40$ then you can readily compute the posterior 
and consider issues of prior sensitivity and predictive performance 
regardless of the choice of prior.  But what if you observe $y=75$? 
Then suddenly you need to be very careful with the choice of prior to 
ensure that your inferences don't blow up.  This doesn't imply that 
the prior should explicitly depend on the measured data, just that a prior that works well in one scenario might be problematic in another.  

Consequently, to ensure a robust analysis we have to go beyond the 
standard Bayesian workflow where the prior distribution is meant to 
be chosen with no reference to the data and, ideally, the data 
generating experiment itself.

\subsection{Existing methods for setting priors already depend on the likelihood}
\label{sec:common_methods}

This tension between the conceptual interpretation of the prior and more 
practical considerations has largely split the long literature of prior 
choice into two sides: either you build a fully subjective prior 
distribution with no knowledge of the  likelihood, or you leverage at 
least some aspects of your likelihood to build your prior.  We refer to 
the first of these positions as \emph{maximalist} in that the prior 
distribution represents, at least ideally, all available information 
about the problem known before the measurement is considered. The 
maximalist prior is implicitly backed up by the Bayesian's willingness 
to bet on it.

Any prior that isn't fully informative but has any sort of theoretical 
or practical benefit leans heavily on some aspect of the likelihood.
The classic example of this is building  priors from the \emph{minimalist}
position which takes data and a model of the measurement process, and considers a prior as little more than an annoying step required to perform a Bayesian analysis.  From this perspective, a 
natural starting point is a noninformative prior.  Although it is 
impossible to define ``noninformative'' with any rigor, the general idea 
is that such a prior affects the information in the likelihood as weakly
as possible. In practice the drive for noninformativity leads to the
naive use of uniform distributions as the limit of an infinitely
diffuse probability distribution.

Related is the idea of the \emph{reference} prior (Bernardo, 1979) 
which, again, serves as a placeholder to allow Bayesian inference to go 
forward with minimal problem-specific assumptions.  These assumptions 
frequently require the statistician to replace knowledge of the likelihood 
with an asymptotic approximation, with the validity of this asymptotic 
regime ultimately affecting the practical performance of the prior.

A \emph{structural} prior encodes mathematical properties such as symmetry 
that represent underlying features of a model.  Examples of structural 
information include exchangeability in hierarchical models and maximum 
entropy models in physics, which Jaynes (1982) and others have applied 
to more general statistical settings.  A structural prior is not 
minimalist as it includes information about the underlying problem which 
is not driven by the measurement process, but neither is it maximalist 
as it does not attempt to include all available information about the
problem at hand.  It also makes the implicit the assumption that the 
structural information is consistent with reasonable data generating
processes.

A \emph{regularizing} prior is designed to yield smoother, more stable 
inferences than would be obtained from maximum likelihood estimation
or Bayesian inference with a flat prior.  Exactly how a regularizing
prior accomplishes this goal clearly depends on the exact nature of
the likelihood itself.  Regularization, even if applied in a Bayesian 
context, is a frequentist goal (Rubin, 1984) in that its success is 
quantified in terms of the statistical properties of the inferences
from an ensemble of possible measurements.

The ground between structural and regularising priors is 
occupied by the more recent, and considerably more hazy, idea of 
\emph{weakly informative} priors that are explicitly designed to encode
information that applies to a general class of problems without 
taking full advantage of problem-specific knowledge 
(Gelman et al., 2008, Simpson et al., 2017).  We argue that these are, 
in a pragmatic sense, \emph{hyper-Jaynesian} in that they are designed 
to regularize inferences with structural information.

\subsection{The role of the prior in generative and predictive modeling}

One aim of this paper is to critically examine the common misconception
that prior modeling doesn't matter.  We argue that for the sorts of 
complex data encountered in modern applications, aspects of the prior 
distributions persist into the posterior and, as such, we have to think 
carefully about how to specify the prior in the context of the likelihood.

In particular, understanding this context requires that we think 
\emph{generatively} by considering the potential measurements consistent
with a given prior and \emph{predictively} by validating those potential
measurements against data that we collect.  The fundamental tool for 
understanding the effect of the prior on inference before data has been
collected is the prior predictive distribution, while the fundamental tool 
for validating the model after data have been collected is the posterior 
predictive distribution.  The careful application of these tools leads us
to some concrete recommendations of how to choose a prior that ensures
robust Bayesian analyses in practice.

\section{A simple motivating example}

To appreciate how impactful the prior can be in a real problem, consider
the paper, ``Beautiful parents have more daughters,'' by Kanazawa (2007), who analyzed data from a longitudinal survey that included a measure of 
adolescent respondents' attractiveness on a 1--5 scale, and followed 
up over the several years and recorded the sex of these people's children.
The sample size was approximately 3000. There was positive correlation between attractiveness and sex ratio in the sample:  a linear regression found
that a 2-point difference in attractiveness corresponds to a 3.0 percentage 
point difference in the probability of a girl birth, with a standard deviation 
of 2.7 percentage points.

This estimate is not statistically significant at the conventional level and, 
as such, would not be generally taken as useful evidence.  The published paper, 
however, featured a comparison between the sex ratio of children of the most 
attractive parents (category 5), compared to those of categories 1--4.  For 
this particular comparison, the proportion of girl births was 8 percentage 
points higher among most attractive parents, and this difference was reported 
as having a $t$-statistic of 2.44, implying a standard error of 3.3 percentage 
points.

For simplicity, we shall proceed with the simple comparison---the estimate of 
$8\pm 3.3$ percentage points---setting aside legitimate concerns of selection 
and multiple comparisons because they are not relevant to our concerns in this 
paper.  The resulting model has two parameters: the probability of girl births 
for beautiful parents, $p_1$, and for others, $p_2$, which we shall parameterize
as $p_2$ and $\delta=p_1-p_2$.  The overall probability of girl births is very 
well estimated from aggregate data, with approximately 4 million births per year 
in the United States, and so here we only  concern ourselves with the 
prior distribution and inference for $\delta$. 

\subsection{Bayesian analysis under different priors}

With a uniform prior on $\delta$, the posterior is proportional to the likelihood, 
approximately normal with mean 0.08 and standard deviation 0.033, thus an
implied 99.2\% chance that beautiful parents are more likely to have girls in the 
general population, and an implied 50\% chance that the difference in probabilities 
exceeds 8 percentage points.

While it is well known that the definition of a uniform prior depends on 
parameterization, in this case the estimated probabilities are so far from 0 and 1 
that there would be essentially no change in posterior inferences if the scale of 
the uniform prior distribution were changed to logit or probit or any other reasonable 
transformation.

The danger here is that {\em any} uniform prior distribution 
contradicts what is known about the stability of the human sex ratio.  Over time and 
across populations, the proportion of girl births has been remarkably stable at about 
48.5\%.  There is some variation but this variation is low, except in cases of 
selective abortion, infanticide, and extreme poverty and famine.  For example, the 
proportion of girl births is about half a percentage point higher among whites than 
blacks in the United States, and there are similar or smaller differences when comparing 
younger mothers and older mothers, babies born in different seasons of the year, and 
other factors that have been studied.  It is hard to imagine sex ratio having a higher 
correlation with parental attractiveness than with these other variables---especially 
given that attractiveness in this particular study was a one-time assessment from a survey 
interviewer.

For a fully informative prior for $\delta$, we might choose normal with mean 0 because 
we see no prior reason to expect the population difference to be positive or negative (see Gelman and Weakliem, 2009, for further discussion of this point) and standard 
deviation 0.001 because we expect any differences in the population to be small, given 
the general stability of sex ratios and the noisiness of the measure of attractiveness.  
The resulting posterior distribution of $\delta$ is then approximately normal with mean 
0.00007 and standard deviation 0.001; that is, our best estimate of the difference in 
sex ratio is 7/1000 of a percentage point, with uncertainty of one-tenth of a percentage 
point.

Somewhere in between these two extremes would be a weakly informative prior, such as normal 
with mean 0 and standard deviation 0.005, which would allow for the population difference 
$\delta$ to be as large as one-half to one percentage point.  The resulting posterior 
distribution is approximately normal with mean 0.002, that is, 0.2 percentage points, 
and standard deviation 0.005, or 0.5 percentage points).  For either the fully informative 
or the weakly informative prior, or variants such as obtained by substituting a $t$ distribution for the normal, the data are so weak that the prior dominates.

\subsection{Understanding the problem}

The point of this example is that for the particular problem of estimating the 
parameter $\delta$---a difference in sex ratios that is certainly
less than 1 percentage point in the general population---the available data from 
3000 survey respondents is laughably weak.  A uniform prior then represents a strong 
statement that $\delta$ can be large, which has has malign consequences for the 
posterior distribution (and for science more generally given that the resulting paper 
was published in a reputable journal and received uncritical publicity in major media,
as discussed by Gelman and Weakliem, 2009).  In other settings, however, a uniform prior 
distribution for a parameter estimated in this way would work just fine: 3000 is a 
large sample size for the purpose of estimating  real underlying differences of 5 percentage points 
or more.  

Thus, the prior distribution here can only be interpreted in the context of the likelihood.   This point is in some sense mathematically 
obvious---after all, the product of any Bayesian inference is the posterior 
distribution which filters the prior through the likelihood---but it contradicts 
the conceptual principle that the prior distribution should convey only information 
that is available before the data have been collected.  The resolution of this apparent contradiction is that priors (and, for that matter, likelihoods) can only be approximate, and the sensitivity of conclusions to certain aspects of the prior will depend on the model for the data.


\section{When exactly is the prior irrelevant in practice?}

The sex ratio example shows how uniform priors can lead to nonsensical 
inferences even with what can seem like a large sample size.  A skeptical reader, however, 
should question whether this example is pathological or indicative 
of a general problem.  

Our experience is that in contemporary statistical practice the problem is indeed general.
The dominance of the information encoded in the measurement depends not only on 
the size of the data but also on the structure of the likelihood and the effect 
being studied.  The more complex the likelihood and the smaller the effect being
considered, the more data are needed to render the prior irrelevant.  To put it another way, when sample sizes are large and data are rich, one can and should be asking more fine-grained questions.
Given the
challenging problems being analyzed at the frontiers of applied statistics,
priors are unlikely to be irrelevant.

\subsection{Uniform priors are not a panacea and can do unbounded damage}

To see why uniform priors are inappropriate, we need to think about what went 
wrong for the sex ratios.  The core of the problem was that, because the true 
difference $\delta$ was small, the data did a bad job of finding it. There is 
an easy frequentist argument for this: the randomness of the data means that 
the maximum likelihood estimator will take $n$ data points and produce an estimate 
of $\delta_\text{true}$ that satisfies $\hat{\delta} = \delta_\text{true} + 
\mathcal{O}_p(n^{-1/2})$.  Unfortunately the random fluctuation in the estimator 
is \emph{additive}, which means that when $\delta_\text{true}$ is small $n$ 
needs to be very large to avoid the natural variation in the data from overwhelming 
the signal.  It's not so much that the uniform prior is inherently bad, but 
rather that its interaction with the likelihood and the data facilitates poor 
performance.  It does not help that uniform or extremely broad prior distributions are often viewed as a safe default prior choice.

\subsection{Asymptotics: so close, yet so far away}

Asymptotic arguments have traditionally played two roles when constructing priors. The first role is to dispel concern by appealing to the Bernstein-von Mises theorem, which suggests that priors have a second-order effect, in the sense that they wash out from the inference faster than the inherent variability of the measurement. The sex ratio example shows that even for quite simple models, this reasoning often doesn't apply in real applications.

The second role asymptotic reasoning has had is in the actual construction of priors. Arguments for the validity of reference priors, maximum entropy priors, and matching priors all rely on some sort of asymptotic justification, which may or may not hold in practice. Indeed, these asymptotic assumptions themselves represent prior information that has been chosen, explicitly or implicitly, to have been included into the model.

The foundational work for most of the priors listed in Section \ref{sec:common_methods},
the key exception being weakly informative priors, was done in the mid-to-late 
20th century.  Critically, these priors were conceived, constructed, and 
publicized long before the computational revolution of the 1990s which has 
driven the the analyses of ever larger and more intricately structured data.  
Ultimately, the applications that these priors were built to solve are not necessarily the same as applications of interest today.

Today, datasets are bigger, and the models needed to capture the structure of those 
data can be considerably more complex than those that were practical in the 
pre-Markov chain Monte Carlo days.  One of the most significant consequences of this big 
data, small signal revolution is the failure of reference priors, maximum entropy 
priors, and most priors that try to match frequentist properties.  The problem 
is that these priors are justified at least partly by asymptotic arguments, 
which requires a strong signal for parameters of interest.
 
In hindsight it is not surprising that using certain priors in situations where 
their asymptotic justification does not hold results in poor statistical analyses.  
Given these challenges, we argue that asymptotic analyses are best limited to 
quantifying when priors may perform poorly rather than motivating the design
of default families of priors.

\subsection{For complex models, certain aspects of the prior will always be relevant}

There are important classes of models in which the number of parameters increases with sample size so that inferences are not identified by data even asymptotically.  In such cases, the posterior eventually concentrates not on a point but rather around some extended submanifold of parameter space---and the projection of the prior along this submanifold continues to impact the posterior 
even as more and more data are collected.  When confronted with such a poorly identified likelihood we must be particularly 
careful to ensure that the prior is sensible along the poorly identified submanifold.

This problem is best understood though another example.  Once the statistical 
equivalent of ``the good china,'' coming out only on special occasions, Gaussian 
processes are now ubiquitous in applied statistics. They also provide a 
tractable case where the interaction between priors, likelihood, and data can be 
laid out precisely.  Consider, a  mean zero GP $x(t)$ defined on the interval $[0,1]$ 
with covariance function
\begin{equation*}
\mbox{cov}(x(t), x(s))
= 
\sigma^2\exp\left( -\kappa |t-s|\right), \qquad \sigma,\kappa \geq 0,
\end{equation*}
where $\sigma$ is the marginal standard deviation and $\kappa$ controls the range of 
the correlation, with $x(t)$ and $x(t\pm 2\kappa^{-1})$ being approximately independent.

The usual asymptotic regime for this sort of model, known as the infill regime, involves 
more observations of the same realisation of the GP within the interval $[0,1]$.  Under 
such infill asymptotics it is well known that the product $\sigma^2 \sqrt{\kappa}$ is 
consistently estimable from the data, while the individual parameters $\kappa$ and 
$\sigma$ are not (Stein, 1999, Zhang, 2004).  The interpretation of this non-identifiability 
is that the data cannot differentiate between a process with a long range and a high 
variance and data with a short range and a low variance. 


While it may be tempting to resolve this non-identifiability by fixing $\kappa$, there 
is strong empirical (Kaufman and Shaby, 2013) and theoretical (van der Vaart and van Zanten, 2009) 
evidence that the models will fit the data better if the parameter is allowed to vary.  
This puts us in an uncomfortable situation.  If there is only a small signal, then the 
problems identified in the sex-ratio example will occur.  On the other hand, even if the 
signal is strong enough to avoid this trap, the prior will still affect the shape of the
posterior along the ridge defined by $\sigma^2 \sqrt{\kappa}= \text{const}$. 

An immediate consequence of this non-identifiability is that a prior on $\kappa$ affects 
the posterior for $\sigma$.  For example, if the prior on $\kappa$ has a very light right 
tail, which penalizes short ranges, then the resulting posterior for $\sigma$ will have almost 
no support around small variances.  Fuglstad et al.\ (2016) recommend specifying the 
prior on $(\kappa, \sigma)$ using coordinates parameterizing motion parallel and orthogonal
to the the ridge.  These more natural coordinates motivated by the structure of the measurement
allow us to specify a meaningful prior on the ridge and ensure the parameter estimates are 
useful.  For example, because we know that the observations are contained within $[0,1]$
we can argue that we want to avoid the part of the ridge where the range is much longer 
than the domain and the marginal variance is compensatingly large.

This Gaussian process example demonstrates that the non-identifiabilities, and near 
non-identifiabilites, of complex models can lead to unexpected amounts of weight being 
given to certain aspects of the prior.  In the Gaussian process case this is fairly 
easy to resolve analytically, but in more complex 
hierarchical models, we know of no general techniques to identify 
problem areas in the parameter space.  In these cases, common sense and even weak subject-matter understanding can translate into useful information along the non-identified submanifold and thus become the basis for effective weakly informative priors.

\section{A prior is more than just a probability measure, so we need to start thinking generatively}

Acknowledging that complex models can be full of hidden pathologies that are 
difficult to explore mathematically is an important first step toward motivating 
robust priors for modern statistical models.  In particular, this realization 
extinguishes the hope of something as mathematically clean as reference or 
maximum entropy priors being useful in practice.  

We have to go further, however, to provide a route toward useful priors and,
specifically, an understanding of why weakly informative priors work so well 
in practice.  If we dig deeper into the reasoning underlying successful weakly 
informative priors, like the half-$t$ on the standard deviation for logistic 
regression (Gelman et al., 2008), then we begin to see a unifying 
principle: those neighborhoods of the parameter space disfavored by a weakly 
informative prior correspond to data generating processes that would look 
strange.  

\subsection{When is a probability distribution a prior?}

This leads to an important but under-appreciated aspect of Bayesian 
analysis.  While every prior distribution is a probability measure, not every 
probability measure is a prior.  A probability measure becomes a prior only 
in the context of a measurement, or, more mathematically, it becomes a prior 
only in the context of a likelihood.  Importantly, we can judge a prior by
examining the data generating processes it favors and disfavors.

A trivial example of this principle is that a probability measure with all 
of its mass on the interval $[-4, -1]$ could never be a prior for the standard 
deviation of a normal distribution as it violates the fundamental non-negative 
nature of a standard deviation.  There is no corresponding data generating 
process.  This is not to say that distributions that are not priors are only 
those that are mathematically precluded.  A probability measure assigning all
of its mass to the single point $\sigma=37$, for example, is unlikely to 
represent any form of reasonable prior belief for this problem.
 
The idea that a probability measure can be precluded from being a prior 
distribution on the grounds that it will not interact sensibly with the 
likelihood to generate a meaningful data generating mechanisms is also 
important in the context of hierarchical models. Consider a logistic 
regression where the logit probability is distributed as 
$\operatorname{logit}(p_i) \sim N(0, \sigma^2)$ and
$p(\sigma)$ is zero for $\sigma\leq 3$ and half-Cauchy for $\sigma>3$.  
This distribution is not mathematically precluded and is not degenerate, but
when paired with the likelihood it yields inferences that concentrate at
the extremes of $p_{i} = 0, 1$.  More formally, the prior ensures that
the model is completely separated no matter the data. 

In this case, we argue that $p(\sigma)$ is not really a prior as the 
induced data generating mechanisms are inconsistent with the data typical
of a logistic regression.  What happens, however, if we actually observe 
data that exhibit complete separation, or, more likely, complete separation 
in one of its subgroups? In that case, we need to make sure that 
whatever prior we use induces data generating processes consistent with
this extreme case.  This is a subtle point; a probability distribution
that serves as a prior for one problem may not serve as a prior for 
another problem.

\subsection{Prior choice is especially important in high dimensions}

A homunculus for the types of pathologies seen in complex models is a 
simple linear regression with Gaussian observation errors with a known 
variance and a design matrix with $p$ orthonormal columns where we measure
only a single datum ($N=1$).  A possible prior for the regression 
coefficients $\mathbf{\beta}$ is independent Gaussians with mean zero
and unit variance; for each $\beta_j$ this prior represents that belief 
that the data are scaled in such a way that the underlying coefficients $\beta_j$ are mostly in the range $(-2,2)$.
The way these independent priors interact with the likelihood, however,
can cause problems when there are many coefficients.  

With the above prior and likelihood, the posterior for $\mathbf{\beta}$ is a product of
independent Gaussians with unit variance and mean given by the 
least squares estimator of $\mathbf{\beta}$.  The problem is that standard 
concentration of measure inequalities show that this posterior is not
uniformly distributed in a unit ball around the ordinary least
squares estimator but rather is exponentially close in the number of
coefficients to a sphere of radius $1$ centered at the estimate.   

That the posterior is very certain that the truth lies somewhere 
near the unit sphere is entirely due to the prior, which strongly 
informs that $\mathbf{\beta}$ lies somewhere near a unit sphere.  If 
this is consistent with your prior knowledge then an iid standard Gaussian 
prior on $\mathbf{\beta}$ is a genuine prior distribution, but if it's not 
then it's just a probability measure in the wrong place at the wrong time.  
This example demonstrates that it's not enough to investigate your prior 
in a parameter-by-parameter manner: it is the \emph{joint} behavior that 
affects inferences and so it is the joint behavior that must be considered.

As we move to more complex and high dimensional problems, subtle joint
behaviors like concentration of measure become more ubiquitous and hence
more critical to consider (Gelman, 1996).  Unfortunately, these behaviors can sneak by 
even seasoned modelers.  You should never underestimate your prior.

\subsection{Sensitivity of the marginal likelihood to the prior}

The guiding principle for prior specification we have emphasized here 
can be encapsulated in the question, Could this prior generate the 
type of data we expect to see?  This accords with the Jaynesian idea 
that a prior should reflect the constraints on the system.  Rather than 
looking for hard constraints which are difficult to elicit for complex 
models, however, we instead focus on ensuring that most of the prior 
mass is in parts of the parameter space that correspond to reasonable
data generating processes.  At the very least we want to ensure that our 
priors don't lead to any unintended structure in the parameter space
and hence in data generated using the full probabilistic model, such as
$\mathbf{\beta}$'s fondness for spheres in the above example.

The idea of building priors that generate reasonable data may seem like 
an unusual idea, but the concept in deeply baked into traditional Bayesian 
practice.  In particular, if we are interested in model selection or model 
averaging, the traditional tool is the marginal likelihood or Bayes factors,
the ratio of marginal likelihoods when comparing two different models 
(Kass and Raftery, 1995).  If we denote our data as $y$ and our parameter vector as
$\theta$, then the marginal likelihood is defined as 
$
p(y) = \int_\Theta p(y | \theta) p(\theta) \,d\theta.
$
More importantly for our considerations here, if we do not evaluate the
marginal likelihood at our data then it becomes a predictive distribution
for the data supported by the prior or the \emph{prior predictive distribution}.
as the predictive distribution.  

From this perspective the marginal likelihood computes the probability of 
a given measurement by first simulating some parameters $\theta \sim p(\theta)$ 
and then simulating a measurement from $p(y \mid \theta)$. The use of Bayes 
factors to choose a model can be seen as an application of decision analysis 
based on an implicit utility function of prior predictive performance, which 
by construction is optimal on average if the model is true.   

If your model is not generative, however, then it makes no sense to compute 
the marginal likelihood as it no longer manifests this interpretation.  If 
you're not worried about building a generative prior, for instance, then it 
would be easy to artificially deflate the marginal likelihood by putting more 
mass on unrealistic parts of the parameter space. Viewed this way, the most 
commonly stated problem with the Bayes factor, that it doesn't make sense when 
using improper prior, is perhaps the least concerning aspect of its application.

In many settings the inappropriateness of the marginal likelihood manifests 
as high sensitivity to aspects of the prior distribution that do not affect
posterior inferences and hence can be difficult to identify.
The problem is that the prior predictive utility function judges models by 
how well they claim to do given their own assumptions and completely ignore
the validity of those assumptions.  In particular, uniform prior distributions 
typically yield atrocious predictive performance.  Given the above considerations,
however, this shouldn't be surprising as the non-generative nature of the uniform
prior obstructs the predictive interpretation of the marginal likelihood.  That
the marginal likelihood can at best be seen as some sort of set-wise limit of 
prior predictive distributions offers little reassurance.  

To some extent these problems disappear when models are assessed using posterior 
predictive distributions rather than prior predictive distributions.
Moving to posterior predictive utility functions, which average over the
data-informed posterior instead of the prior, is more robust from this perspective.
Not that it's universally correct, just more robust if your goal is predictive 
performance.

As with many things, the truth here is conditional.  Applications of posterior 
predictive distributions are robust to prior specification only when the
details of the prior are washed out by the likelihood.  In the example of the 
previous section, as with many contemporary problems, this was not true. In 
these cases we need to use more principled priors, such as weakly informative 
priors to get a posterior distribution, and hence a marginal likelihood, that
is sensible.  In particular, model selection through posterior predictive is 
relatively stable under weakly informative priors.

Posterior predictive selection is stable under weakly informative priors, 
but does this mean that marginal likelihoods are stable?  Unfortunately, the 
answer in general is no.  The posterior often does not well identify the prior: 
many priors will yield the same posterior given a common likelihood.  This 
means that recommendations for diffuse and even weakly informative priors are 
not well suited to applications of marginal likelihoods.  If you want to use 
marginal likelihoods then you had better be willing to defend every detail
of your prior, even those that might seem otherwise irrelevant.  For a 
parameter of unit scale with an informative likelihood and zero-mean prior,
for example, a change in the prior standard deviation from 100 to 1000 leads 
to an approximate drop in a factor of 10 in the marginal likelihood, even 
while having no appreciable effect on the posterior distribution and
corresponding inferences.

\section{Generative priors need to be prediction focused}

In the previous section, we argued that Bayesian model comparison is at
least implicitly a statement about how well a model predicts new measurements.  
These comparisons fall into two varieties: prior predictive methods, i.e. 
marginal likelihoods, where you try to predict measurements using the prior
uninformed by any data, and posterior predictive methods where you try to 
predict new measurements using a posterior informed from previous measurements.  
Of these options, posterior predictive methods offer meaningful, robust model 
selection procedures, while prior predictive methods can give meaningless 
results, especially for prior models that can't be viewed generatively.  

This suggests that as well as being generative, we should ensure our priors 
facilitate good predictions. These are not the same thing.  A prior is 
generative if the prior predictive distribution generates only data deemed
consistent with our understanding of the problem.  On the other hand, a prior 
has good predictive performance if the posterior predictive distribution is
consistent with the true data generating process and can predict new data 
generated from similar experiments.  Importantly, a good predictive prior 
allows the corresponding model to generalize and avoid overfitting.

\subsection{In the sea of complex models, the leviathan is  overfitting}

One of the common ways that complex models fail to produce good posterior 
predictive distributions is when the model ``overfits.''  Although useful 
as a concept, the definition of overfitting is difficult to pin down, in part because the concept is generally understood as a comparison between fits to training and test data, but the bare-bones Bayesian formulation $p(\theta|y)\propto p(\theta)p(y|\theta)$ makes no mention of test data.  So to even consider overfitting it is necessary to consider some partitionable structure of data.  Alternatively we can say that a complex model overfits when it contains a 
simpler submodel that does a better job at predicting new measurements---but then this requires some idea of workflow or network of models, as, again, there is no concept of ``submodel'' in the most basic expression of Bayes' theorem.
To our knowledge, the penalized complexity prior framework of Simpson 
et al.\ (2017) was the first place that the avoidance of overfitting was explicitly linked to prior construction.    Their big idea was that, for a complex model 
$M_\Theta$, the simpler model that potentially generalizes better can be 
written as $M_{\theta_0}$, where $\theta_0 \in \Theta_{0} \subset \Theta$ 
is one of a finite set of parameter vectors that describe simpler sub-models 
of $M_\Theta$.  

With this structure in place, it's possible to talk about the \emph{a priori} 
probability that $M_\Theta$ overfits, that is the prior probability that 
$\theta$ is not in some small neighborhood of $\Theta_0$.  That is, you can 
talk about overfitting before you make a measurement by talking about how 
often draws from the prior distribution give values sufficiently far away 
from $\Theta_0$.  This allows you to check for potential overfitting before 
a data analysis and then check for actual overfitting, using posterior 
predictive checks, after the model has been fit to the data.

\subsection{Overfitting leads to poor posterior predictive performance} 

Another way of understanding why priors that put sufficient mass around 
the simpler sub-models can give better performance is by reconsidering 
Stein's  shrinkage estimator for the mean of a normal distribution.  
If you see one data point $y \sim N(\mu,\sigma^2 \mathbf{I}_N)$ where 
$\sigma$ is known, $\mu$ is unknown, and $N$ is the dimension of the 
observation, then the best estimator of $\mu$, in the sense that it's 
equivariant and minimax, is $\hat{\mu}(y) = y$.  Stein's example showed 
that this ``best'' estimator can \emph{always} be improved in the sense 
that we can find a new estimator $\tilde{\mu}(y)$ such that 
$\| \mu - \tilde{\mu}(y)\| \leq \| \mu - \hat{\mu}(y)\| $ whenever 
$N\geq 3$, where the inequality is often strictly true.  Stein's trick 
was to notice that the point $\mu=0$  has the property that if $y$ is 
sufficiently close to it, it's better from an $\ell_2$ error point of 
view to estimate $|\mu| \ll y$ than to estimate $\mu=y$.  A now standard 
analysis shows that this is true if $ |y| \leq \sqrt{m}\sigma$ and 
Stein's estimator corresponds to a prior that puts a lot of \emph{a priori}
probability mass on this ball.

A big barrier to extending this idea into more practical situations 
is finding the equivalent concept of a ``ball of radius $\sqrt{m}\sigma$''.  
A moment thinking about the structure of the problems suggests that the 
Euclidian nature of the ball is mostly an artifact of the problem, rather 
than a generalizable quantity.  It is an open question as to what shape 
these balls should have in general, i.e. what shape should a neighborhood
around $\Theta_0$ have, related to the fact that all of the tools that 
we use to analyze these types of properties eventually require us to 
marginalize out the sort of parameters that can only do in very simple 
cases such as when the problem is estimating the mean of a multivariate 
normal distribution. 

Simpson et al.\ (2017) made a heuristic argument that if the model 
is parameterized so that each parameter controls a different aspect of 
the complex model then it's sufficient to consider just the local shape 
of the parameter space.  They did this using a localized version of the 
Kullback-Leibler divergence, which seems to work reasonably well in the 
cases that have been examined (Klein and Kneib, 2016).  There is some 
immediate work, however, that needs to be done to extend this to more 
general classes of models.

\subsection{Don't forget your roots: predictive priors aren't always generative}

While this section has focused on outlining methods that build priors 
that give good predictive properties, this still doesn't absolve us of
our obligation to ensure that the resulting prior is generative. 
Importantly, these can be competing aims.  

Consider, for instance, the model $y |\mu \sim N(\mu,1)$, 
$\mu |\sigma \sim N(0,\sigma^2)$, $\sigma \sim p(\sigma)$.  The arguments 
of Simpson et al.\ (2017), which are inspired greatly by Gelman (2006) 
and Gelman et al.\ (2008), argue that $p(\sigma)$ should have finite, 
non-zero density at $\sigma=0$. On the other end, Polson and Scott (2012) 
argue persuasively that, from an admissibility point of view, the prior on 
$\sigma$ should have very heavy tails, eventually advocating the half-Cauchy 
prior on $\sigma$ advocated originally by Gelman (2006).  

We argue that this prior is usually not generative.  Around $1.2\%$ of the 
time a half-Cauchy prior with unit scale parameter will propose a standard 
deviation of more than $50$, which seems unrealistic if the initial parameters 
of the model are reasonably scaled.  This is also born out in numerical 
pathologies described by Piironen and Vehtari (2015).  

We currently do not have a good recommendation on how heavy the tails of this 
parameter should be.  Experimentally, however, we know that if you're confident 
of your scaling then a half-normal on the standard deviation works well. If 
you're less confident then an exponential is effective, and if you're even less  
confident then a half-Student-t with more than $3$ degrees of freedom is useful. 
And if you're really struggling then the half-Cauchy is always there for you. 

The practical guidance is to remember that you cannot have too many parameters 
with a heavy tail in the model, lest the joint prior put too much probability
mass onto a bad part of the parameter space. Our rule of thumb is that the 
heavier the tail on one component of the model, the less ``ambitious'' you can 
be with the rest of the model. 

\section{Discussion}

The literature on the choice of Bayesian priors is mixed when it comes to 
the likelihood function.  On one hand, the mathematics of Bayesian inference 
and the very term ``prior'' suggest that the model $p(\theta)$ should depend 
only on the space of $\theta$ and its context within the application, not on 
any hypothetical data that might come later.  On the other hand, Jeffreys' 
prior (Jeffreys, 1961, Kass and Wasserman, 1996), long a popular default, is 
explicitly defined in terms of the likelihood function, and other 
suggestions such as fractional or intrinsic Bayes factors 
(O'Hagan, 1995, Berger and Pericchi, 1996) additionally depend on the concept 
of individual data points.  For problems with unbounded parameters, the
utility of extremely diffuse and even uniform prior distributions require 
conditions on the likelihood such that the posterior will be proper.  This 
can lead to awkward compromises where the prior is augmented a posteriori
to account for certain data patterns such as separation in logistic regression.

Improper priors are inimical to coherent Bayesian inference, but for many 
problems the structure of parameter space is such that any prior that 
respects certain natural symmetry principles will be improper.  For a 
serious Bayesian this implies that these symmetry properties can be
insufficient or even entirely inappropriate.  For example, it would not make 
sense to model the prior for the probability of a girl birth as being 
translation-invariant on the logit probability scale.  In settings with
strongly informative data and sparse prior information, such concerns 
can be safely ignored.

We view much of the recent history of Bayesian inference as a set of 
converging messages from many directions---theoretical, computational, 
and applied---all pointing toward the benefits of including real, 
subject-matter-specific, prior information in order to get more stable 
and accurate inferences.  This puts new and significant burdens on the
developers and users of Bayesian methods, and an obligation for 
statisticians to develop default priors, or more generally procedures 
for researchers to build bespoke priors, going beyond the traditional 
recommendations.  At the same time, those researchers need to recognize
the importance of the prior and spend the time encoding their expertise
in probabilistic form.  

In this paper we have argued that a prior can in general only be 
interpreted in the context of the likelihood with which it will be paired.  
This pairing is best understood through the context of prediction and 
the properties of the posterior predictive distribution which quantify
how appropriate a prior might be for a particular problem.  This observation 
is critical to methodologists and practitioners in guiding their efforts 
toward default and subject-matter-specific prior distributions (Stan Development Team, 2017). 

\section*{References}

\noindent

\bibitem Berger, J. O., and Pericchi, L. R. (1996).  The intrinsic Bayes factor for model selection and prediction. {\em
Journal of the American Statistical Association} {\bf 91}, 109--122.

\bibitem Bernardo, J. M. (1979).  Reference posterior distributions for Bayesian inference (with discussion).  {\em Journal of the Royal Statistical Society B} {\bf 41}, 113--147.

\bibitem Fuglstad, G. A.,  Simpson, D.,   Lindgren, F., and  Rue, H. (2017). Constructing priors that penalize the complexity of Gaussian random fields. \url{http://arXiv:1503.00256}.

\bibitem Gelman, A. (1996).  Bayesian model-building by pure thought:  Some principles and
examples.  {\em Statistica Sinica} {\bf 6}, 215--232. 

\bibitem Gelman, A. (2006). Prior distributions for variance parameters in hierarchical models. {\em Bayesian Analysis} {\bf 1}, 515--534.

\bibitem Gelman, A., Carlin, J. B., Stern, H. S., Dunson, D. B., Vehtari, A. and Rubin, D. B. (2003). {\em Bayesian Data Analysis}, third edition. London:  CRC Press.

\bibitem Gelman, A., and Hennig, C. (2017).  Beyond subjective and objective in statistics (with discussion). {\em Journal of the Royal Statistical Society}.

\bibitem Gelman, A., Jakulin, A., Pittau, M. G., and Su, Y. S. (2008).  A weakly informative default prior distribution for logistic and other regression models. {\em Annals of Applied Statistics} {\bf 2}, 1360--1383.

\bibitem Gelman, A., and Weakliem, D. (2009).  Of beauty, sex, and power: Statistical challenges in estimating small effects. {\em American Scientist} {\bf 97}, 310--316.

\bibitem Jaynes, E. T. (1982).  On the rationale of maximum-entropy methods.
{\em Proceedings of the IEEE} {\bf 70}, 939--952.

\bibitem Jeffreys, H. (1961).  {\em Theory of Probability},
third edition.  Oxford University Press.

\bibitem Kanazawa, S. (2007).  Beautiful parents have more daughters: A further implication of the generalized Trivers–Willard hypothesis (gTWH).  {\em Journal of Theoretical Biology} {\bf 244}, 133--140.

\bibitem Kass, R. E., and Raftery, A. E. (1995).  Bayes factors and
model uncertainty.  {\em Journal of the American Statistical Association}
{\bf 90}, 773--795.

\bibitem Kass, R. E., and Wasserman, L. (1996).  The selection of prior distributions by formal rules.  {\em Journal of the American Statistical Association} {\bf 91}, 1343--1370. 

\bibitem Kaufman, C. G., and  Shaby, B. A.  (2013). The role of the range parameter for estimation and prediction in geostatistics. {\em Biometrika}  {\bf 100}, 473--484.

\bibitem Klein, N., and Kneib, T. (2016). Scale-dependent priors for variance parameters in structured additive distributional regression. {\em Bayesian Analysis} {\bf 11}, 1071--1106.

\bibitem O'Hagan, A. (1995).  Fractional Bayes factors for model comparison
(with discussion).  {\em Journal of the Royal Statistical Society B} {\bf 57},
99--138.

\bibitem Piironen, J. and Vehtari, A. (2015). Projection predictive variable selection using Stan+ R.
\url{http://arXiv:1508.02502}.

\bibitem Polson, N. G. and Scott, J. G. (2012). On the half-Cauchy prior for a global scale parameter.
{\em Bayesian Analysis} {\bf 7}, 887--902.

\bibitem Rubin, D. B. (1984).  Bayesianly justifiable and relevant frequency calculations for the applied statistician.  {\em Annals of Statistics} {\bf 12}, 1151--1172.

\bibitem Simpson, D., Rue, H., Riebler, A., Martins, T. G., and Sorbye, S. H. (2017).  Penalising model component complexity: A principled, practical approach to constructing priors.  {\em Statistical Science} {\bf 32}, 1--28.

\bibitem Stan Development Team (2017).  Prior choice recommendations.  \url{https://github.com/stan-dev/stan/wiki/Prior-Choice-Recommendations}

\bibitem Stein, M. L. (1999).  {\em Interpolation of Spatial Data: Some Theory for Kriging}. Springer.

\bibitem van der Vaart, A. W., and van Zanten, J. H. (2009). Adaptive Bayesian estimation using a Gaussian random field with inverse gamma bandwidth.  {\em Annals of  Statistics} {\bf 37}, 2655--2675.

\bibitem Zhang, H. (2004). Inconsistent estimation and asymptotically equal interpolations in model-based geostatistics.  {\em Journal of the American Statistical Association} {\bf 99}, 250--261.

\end{document}